\documentclass[11pt]{article}
\usepackage{coling2018}
\usepackage{times}
\usepackage{url}
\usepackage{latexsym}
\usepackage{graphicx}
\usepackage{float}
\usepackage{threeparttable}
\usepackage{amsmath}
\usepackage{amssymb}

\title{Exploiting local and global performance of candidate systems for aggregation of summarization techniques}

\author{Parth Mehta \\
	IR\&LP Lab \\
	DA-IICT, India\\
  {\tt parth\_me@daiict.ac.in} \\\And
  Prasenjit Majumder \\
	IR\&LP Lab \\
	DA-IICT, India\\
	{\tt p\_majumder@daiict.ac.in} \\}

\date{}
\begin{document}
	\maketitle
	\begin{abstract}
		With an ever growing number of extractive summarization techniques being proposed, there is less clarity then ever about how good each system is compared to the rest. Several studies highlight the variance in performance of these systems with change in datasets or even across documents within the same corpus. An effective way to counter this variance and to make the systems more robust could be to use inputs from multiple systems when generating a summary. In the present work, we define a novel way of creating such ensemble by exploiting similarity between the content of candidate summaries to estimate their reliability. We define GlobalRank which captures the performance of a candidate system on an overall corpus and LocalRank which estimates its performance on a given document cluster. We then use these two scores to assign a weight to each individual systems, which is then used to generate the new aggregate ranking. Experiments on DUC2003 and DUC 2004 datasets show a significant improvement in terms of ROUGE score, over existing sate-of-art techniques.
		
	\end{abstract}
	
	\section{Introduction}
	An ever increasing interest in the field of automatic summarization has led to a plethora of extractive techniques, with a lack of clarity about which is the best technique. Unclear implementation details and variation in evaluation setups only make the problem worse. However, there is no doubt that none of these techniques would always work. Not only will there be a difference in performance across different datasets, there is also a good amount of variation across documents in the same dataset. In several cases this difference is also attributed to the fact that different ROUGE setups are used for evaluation, which can result substantial variation in the scores. \cite{hong2014repository} propose using a fixed set of parameters for ROUGE and report comparable results for several summarization algorithms. Even with this normalization, the system performance still varies a lot, and there is a possibility of exploiting this variation to generate better performing systems. To give an example, we show a simple comparison of two extractive summarization systems from those used by \cite{hong2014repository} in their experiments. We pick two extreme systems, in terms of performance, from the those reported in the work. The FreqSum system\cite{nenkova2006compositional}, which has the weakest performance, was compared to the DPP system\cite{kulesza2012determinantal} which was the best performing system amongst those compared. On DUC 2004 dataset, FreqSum performed better on more than 10$\%$ of the document clusters. There are documents for which a system which is overall very weak, outperforms the system that has a very good performance on an average. Going a step further, an oracle of just the five baseline systems, outperforms DPP in a little over fifty percent of the document clusters. The argument is clear: ensemble of several systems can definitely improve the performance as compared to individual systems. The ideal way of forming an ensemble summary would be to select the relevant information from each candidate while discarding the rest.\\ 
	
	In this work we propose a new method for estimating the authority of a particular system for a given document and at the same time also estimating the importance of each sentence within the summary generated by that system. The ensemble summary is then a function of the authority of each candidate system as well as the relative importance of each sentence in the candidate summaries. The fact, that informative or \emph{summary worthy} sentences in a document cluster are much less in number compared to the non-informative ones, forms the basis of our hypothesis. We argue that since this content is much less, any substantial overlap between two summaries will likely be due to the \emph{important} content rather than the redundant one. Simply because there is less content to choose from when it comes to the important sentences two good summaries will have a good amount of overlap in content. At the same time it is highly unlikely that two summaries will also chose the same not-important content. Keeping this argument in mind we associate higher similarity in content between two summaries with the summaries having more informative content. \\
	
	We use graph based ranking that takes into account the similarity of a candidate summary with other candidates to generate its local (or document specific) ranking. We also determine the overall global ranking of a system from its ROUGE score on a development dataset. In the same way \emph{informativeness} of a sentence is linked to its overlap with sentences of other summaries. The HybridRank model proposed here, combines these three factors to generate a new aggregate ranking of sentences.
	
	\section{Related Work}
	
	In contrast to the amount of attention automatic summarization, and especially the extractive techniques, has achieved from researchers, aggregation techniques have been explored little. This is counter intuitive given several studies which show that even in cases where two systems achieve a comparable ROUGE score, the actual content can be quite different \cite{hong2014repository}. Existing aggregation techniques can be broadly classified into two categories: \emph{rank aggregation} and \emph{summary aggregation}. These techniques are used post-summarization, i.e. each candidate system is first used without any modification. The output, whether ranked lists or summaries, are then used for aggregation. The former method solely relies on candidate summaries, without any information or assumption about the original sentence rankings. In contrast the latter combines existing ranked lists to generate a new aggregate ranking. Apart from these two, there is another type of aggregation which combines various aspects of candidates and incorporates them into the algorithm itself.\\
	
	\cite{pei2012supervised} and \cite{hongsystem} are two instances of the summary aggregation techniques. \cite{pei2012supervised} use SVM-Rank to learn the optimum ranking of sentences from candidate systems. Each sentence is labeled as $-1,0,1$ depending on its \emph{summary worthiness} and then it is used to learn pairwise ranking for each sentence. As opposed to this, \cite{hongsystem} attempt to generate rankings of entire summaries to find out the combination of sentences that maximizes ROUGE-1 or ROUGE-2. They use summaries from four different systems to begin with. Next they combine these summaries and list out all possible candidate summaries by selecting a fixed number of sentences from the combined set. Several word and summary level features are then used to train a SVM-Rank algorithm that can learn to rank candidate summaries to maximize ROUGE scores. \\
	
	Excluding the common techniques like \emph{Round Robin}, \emph{Borda Count} or \emph{Reciprocal Rank} only notable attempt at using rank aggregation was made by \cite{wang2012weighted}. The \emph{weighted consensus summarization} system proposed by them treats rank aggregation as a optimization problem. This approach also introduced the concept of consensus between candidate summaries. They create a weighted combination of ranked lists, under the constraint that the aggregate ranking be as close to original rankings as possible. Unlike the approach proposed in this paper, \cite{wang2012weighted} do not differentiate candidates based on their \emph{trustworthiness}, and instead try to maintain as much information from each candidate as possible. Apart from not taking into account the content of candidate summaries, there are two major limitations with this approach. One, it uses $L_1$ norm for computing similarity (or distance) between candidate rankings, which can be a sub-optimal choice when compared to traditional metrics like \emph{Kendall's Tau}. The other major problem is, this technique tries to optimize rankings over all sentences in the document cluster. So even if the systems agree on ranking of top-k sentence, which actually go into the summary, the aggregation will try to optimize over the entire rankings. This is not only unnecessary, but can also affect the performance adversely.\\
	
	Neither \emph{summary aggregation} nor \emph{rank aggregation} take into consideration the original summarization algorithms or in any way modify them. There have been a few attempts at modifying summarization algorithms to incorporate features from several candidates into a single algorithm. But such attempts are limited by the non-triviality of being able to meaningfully combine unique aspects of the candidates. One popular approach of this kind is combining several sentence similarity scores to generate an aggregate similarity which can then be used by any of the existing algorithms. The \emph{MultSum} technique proposed by \cite{mogrenextractive} builds upon the submodular optimization technique\cite{lin2012learning}. They replace the cosine similarity used in the original approach with multiplicative combination of several sentence similarity measures. This can be extended to several techniques which rely on a sentence similarity metric. But in general it is difficult to define an \emph{aggregate} for other components of a summarization algortihm. As an alternate, \cite{mehta2018effective} propose combining several aspects of candidate systems like \emph{sentence similarity}, \emph{ranking algorithm} and \emph{text representation scheme} in a post-summarization setup. This approach falls within the rank aggregation technique, except that the candidate systems are created by varying one of these three components of the original systems. The \emph{LexRank} algorithm can be used with Kullback Leibler Divergence or word overlap, instead of cosine similarity. Such variations are then combined using existing rank aggregation techniques.

	\section{Proposed Approach}
	In this paper we propose a new summary aggregation technique which takes into account the content of each candidate systems, rather than only ranked lists as done in \cite{wang2012weighted}. In a multi document summarization setup, especially in case of newswire, it is quite common to have duplicate or near duplicate content getting repeated across multiple documents. To put things into perspective, DUC 2003 dataset has, on an average, 34 sentences per cluster which have an exact match within the document cluster. More than 50 sentences have a 80\% match with another sentence. Most existing summarization techniques do not handle this redundancy explicitly, neither do most existing aggregation techniques. This has a huge impact on a class of aggregation techniques which rely only on ranked lists aggregation without taking into account the actual content of individual summaries. For instance, consider the example shown below where $S_1$ and $S_2$ are individual sentence rankings and $S_A$ is the aggregate ranking. This would be fine if each sentence is different and equally important. But consider a case where $sim(s_{1}, s_4) = 1$. The fact that $s_1$ and $s_5$ are repeated across documents makes them more important. But the rank aggregation techniques fail to take into account their actual content and treat these separately, which may result in lowering of their aggregate scores. $S_A^*$ indicates the ideal rank aggregate. 
	\begin{figure}[H]
		\centering
		\includegraphics[scale=0.6]{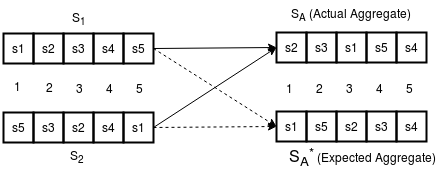}
		\caption{Issue with rank aggregation}
	\end{figure}
	Instead, in the proposed approach we take into account content of the summaries being aggregated to assign weights to candidate systems. The proposed system takes into account the overlap in content between summaries of the candidate systems to compute their reliability, and in turn use it to assign relative importance to individual sentences in those summaries. \\
	
	 The proposed method estimates importance of a given sentence in a particular candidate summary, and then uses this information to generate a meta-summary. In contrast to the solution proposed by \cite{hongsystem}, where they use word and summary level features to estimate the relative importance of each candidate, we propose using similarity between candidates as a measure for their importance. We argue that given two good summaries there is bound to be a good amount of overlap in their content, simply because the relevant content in a document cluster is much less compared to non-relevant content. Arguing the other way round, if two summaries have a good amount of overlap, it is likely because both have more \emph{informative} content. Due to a large number of \emph{non-informative} sentences, selecting the same set of \emph{non-informative} sentences in two distinct summaries is much less likely. The only possible exception where two bad summaries might have higher overlap, will be when both have very similar sentence rankings. As the number of candidate systems increase, this will automatically be taken care of. We start with a simple method to estimate informativeness of sentences in candidate summaries.
	
	\subsection{SentRank}
	This system takes into account similarity of each sentence in a candidate summary, with that of other candidate summaries, and uses it to assign a relative importance to the sentence. The argument presented above, in favor of using similarity as a measure for reliability of a summary can easily be extended at sentence level. A sentence that is \emph{summary worthy} will share more information with another \emph{summary worthy sentences}.
	
	For $i^{th}$ sentence in the $j^{th}$ candidate summary($s_{ij}$), we find the best matching sentence in the remaining candidate summaries. The score of that sentence can then be computed as shown in equation \ref{eq1} below. Score of the sentence is the sum of its similarity with the best matching sentences from remaining candidates. Here $j$,$k$ are the candidate systems. $i$ and $l$ are the sentences in candidate $J$ and $K$ respectively.
	
	\begin{eqnarray}
	\boldsymbol{R}(s_{ij})  = \sum_{k, k\neq j} \max_{l}(Sim(s_{ij}, s_{lk}))
	\label{eq1}
	\end{eqnarray}
	
	Each sentence in the candidate summary is then ranked according to their score $\boldsymbol{R}$ and top k sentences are selected in the summary. We experimented with n-gram overlap, cosine and KL Divergence for computing the similarity between sentences and empirically select cosine similarity, which is also used in the subsequent systems.
	
	\subsection{GlobalRank}
	One limitation of the SentRank approach proposed above is that does not take into account the reliability of candidate systems into account and treats each candidate equally. A sentence that comes from a well performing candidate system is more likely to be \emph{informative} compared to a sentence from a poor summary. The proposed \emph{GlobalRank} system does exactly that. It builds over the SentRank system by incorporating a candidates \emph{global reputation} score into the sentence ranking scheme. The new scoring mechanism is shown in equation \ref{eq2} below. $G(k)$ refers to the global reputation of candidate system $k$.
	
	\begin{eqnarray}
	\boldsymbol{R}(s_{ij})  = \sum_{k, k\neq j} G(k)*\max_{l}(Sim(s_{ij}, s_{lk}))
	\label{eq2}
	\end{eqnarray}
	
	$G(k)$ is estimated using the average ROUGE-1 recall of each candidate systems as shown in equation 4 and 5 below. $R1_k$ is the rouge-1 recall of the $k^{th}$ candidate. $R1^{'}$ is the normalized version of $R1$. Here we do not subtract mean, to avoid negative values in scoring, and instead subtract the minimum of $R1^{'}$. Additionally we scale it using a scaling factor $a$, which is dependent on the total number of candidate systems. We empirically set $a$ to $0.1$. We used the results on DUC2002 dataset for estimating the ROUGE-1 recall, and in turn the GlobalRank of a candidate.
	
	\begin{eqnarray}
	G(k) = aR1'(k) \\
	R1^{'}_{k} = \frac{R1_k}{\sigma(R1_k)} - \min_{k}\Big[\frac{R1_k}{\sigma(R1_k)}\Big] 
	\label{eq3}
	\end{eqnarray}
	
	As compared to SentRank, which can be overwhelmed by too many poor performing systems, GlobalRank provides a smoothing effect, by giving more importance to the systems that are known to perform well generally.  
	
	\subsection{LocalRank}
	One major limitation of the existing aggregation systems, which we highlighted in section 2,  is their inability to predict which candidate system will perform better for a given document cluster. Neither of the systems suggested above, \emph{SentRank} and \emph{GlobalRank}, address this problem. The next system, \emph{LocalRank} tries to mitigate this problem. We do not rely on any lexical or corpus specific features, simply because the training data is not sufficient to estimate these features reliably. Instead we continue on our line of argument, using the similarity between summaries as a measure of reliability. For a give document cluster, we estimate reliability of a candidate $k$ from the content it shares with other candidates, and also the reliability of those candidates.  We first create a graph with the nodes as the candidate summaries and edge as the similarity between nodes. Each candidate starts with the same reputation score or LocalRank ($L$). The local rank is then updated iteratively using the pagerank algorithm \cite{page1999pagerank}. The Local rank for a given node is estimated as shown in equation \ref{eq4} below:
	
	\begin{equation}
	L(k) = \sum_{j}L(j)*Sim(S_j, S_k)
	\label{eq4}
	\end{equation}
	
	$L(k)$ indicates local rank of $k^{th}$ candidate, $S_k$ indicate summary generated by the $k^{th}$ candidate. We use cosine similarity as the similarity score. The overall sentence scores are then computed just like in the GlobalRank algorithm.
	
	\begin{eqnarray}
	\boldsymbol{R}(s_{ij})  = \sum_{k, k\neq j} L(k)*\max_{l}(Sim(s_{ij}, s_{lk}))
	\label{eq5}
	\end{eqnarray}
	
	\subsection{HybridRank}
	
	While LocalRank is useful for estimating how well a given candidate might perform for a given document cluster, it does not make use of the actual system performance. HybridRank overcomes that limitation. As the name suggests, HybridRank combines strengths of both GlobalRank as well as LocalRank by taking a weighted combination of both. The HybridRank is defined in the equation \ref{eq6} below
	
	\begin{equation}
	H(k) = \alpha L(k) + (1-\alpha)G(k)
	\label{eq6}
	\end{equation}
	
	Here the value of $\alpha$ determines the balance between Local and GlobalRank. Higher value of Alpha gives more importance to the estimate of how good a system will perform on a particular cluster, while ignoring the overall aggregate performance of candidate. $\alpha = 0$ leads to the original GlobalRank, without any local information. We empirically set the value of $\alpha$ to 0.3. Once the systems are ranked, the sentence rankings are computed in the same manner as LocalRank or GlobalRank (equation \ref{eq2} and \ref{eq5}).
	
	\section{Experimental Setup and Results}
	
	We report the experimental results on the DUC 2003 and DUC 2004 datasets. We report standard ROUGE scores\cite{lin2004rouge} that are well accepted across the community, ROUGE-1, ROUGE-2 and ROUGE-4 recall. To be consistent and in order to make our work reproducible, we use the same set of ROUGE parameters as that used by \cite{hong2014repository}\footnote{ROUGE-1.5.5.pl -n 4 -m -a -x -l 100 -c 95 -r 1000 -f A -p 0.5 -t 0}. We use eleven candidate systems which are a mix of several state of art extractive techniques and other well known baseline systems. The complete list is shown below, with a brief description of each system. For the DUC 2004 dataset, \cite{hong2014repository} provide summaries for all these systems. We directly use these pre-generated summaries provided for that particular year, to provide a fair comparison. We generated results for the DUC 2003 dataset ourselves. We do post-processing on the generated summaries, by dropping sentences that have a cosine similarity of more than 0.5 with the already selected sentences. Apart from that no other pre/post-processing was done.
	
	We use the following systems as candidate systems for our experiments: 
	\begin{itemize}
		\item[] \textbf{LexRank} This method proposed by \cite{erkan2004lexrank} treats each document as a undirected graph and each sentence in the document constitutes a node. The edges represent cosine similarity between nodes. The importance of each node is then iteratively determined by the number of other nodes to which it is connected and the importance of those nodes.
		\item[] \textbf{FreqSum} A simple approach\cite{nenkova2006compositional} that ranks each sentence based on the frequency of its constituent words. Higher the frequency more the informativeness of the word.   
		\item[] \textbf{TsSum} This method defines \emph{topic signatures} as the words which are more frequent in a given document compared to a background corpus\cite{lin2000automated}. Sentenced with more \emph{topic signatures} are considered more informative
		\item[] \textbf{Greedy-KL} This method follows a greedy algorithm\cite{haghighi2009exploring} to minimize the KL Divergence between the original document and resultant summary. Sentences are sequentially added to the summary so as to minimize the KL-Divergence between word distributions of the set of sentences selected so far and the overall document
		\item[] \textbf{CLASSY04} Judged best among the submissions at DUC 2004\cite{conroy2004left}, uses a hidden markov model with topic signatures as the features. It links the usefulness of a sentence to that of its neighboring sentences.
		\item[] \textbf{CLASSY11} This method builds over the CLASSY04 technique, and uses topic signatures as features while estimating the probability that a bigram will occur in human generated summary. It employs non negative matrix factorization to select a subset of non-redundant sentences with highest scores. 
		\item[] \textbf{Submodular} \cite{lin2012learning} treat summarization as a submodular maximization problem. It incrementally computes the \emph{informativeness} of a summary and also provides a confidence score as to how close the approximation is to globally optimum summary.
		\item[] \textbf{DPP} Detrimental point processing\cite{kulesza2012determinantal} is the best performing state-of-art system amongst all the candidate systems. DPP scores each sentence individually, while at the same time trying to maintain a global diversity to reduce redundancy in the content selected.
		\item[] \textbf{RegSum} uses diverse features like parts of speech tags, name entity tags, locations and categories for supervised prediction of word importance\cite{hong2014improving}. The sentence with most number of \emph{important} words are then included in the summary.
		\item[] \textbf{OCCAMS\_V} The system by \cite{davis2012occams}, employs LSA to estimate word importance and then use the budgeted
		maximal coverageand the knapsack problem to generate sentence rankings.
		\item[] \textbf{ICSISumm} treats summarization as a global linear optimization problem\cite{gillick2009icsi}, to find globally best summary instead of selecting sentences greedily. The final summary includes most important concepts in the documents.
	\end{itemize}
	
 As benchmark ensemble techniques, we use two existing techniques: \emph{Borda count} and \emph{Weighted consensus summarization}\cite{wang2012weighted} described in section 2. For the GlobalRank system, we used DUC 2002 dataset as a development dataset to estimate the overall performance of candidate systems. We ranked the systems based on ROUGE-1 recall scores for this purpose. The results are shown in table 1 below:
	

	\vspace{3mm}
	
	\begin{table*}[t]
		\centering
		\begin{tabular}{|c|c|c|c| c |c | c|}
			\hline
			            &                           \multicolumn{3}{c|}{DUC2003}                            &                           \multicolumn{3}{c|}{DUC2004}                            \\ \hline
			  System    &            R-1            &            R-2            &            R-4            & R-1                       & R-2                       & R-3                       \\ \hline
			  LexRank   &          0.3572           &          0.0742           &          0.0079           & 0.3595                    & 0.0747                    & 0.0082                    \\
			  FreqSum   &          0.3542           &          0.0815           &          0.0101           & 0.3530                    & 0.0811                    & 0.0099                    \\
			   TsSum    &          0.3589           &          0.0863           &          0.0103           & 0.3588                    & 0.0815                    & 0.0103                    \\
			 Greedy-KL  &          0.3692           &          0.0880           &          0.0129           & 0.3780                    & 0.0853                    & 0.0126                    \\
			 CLASSY04   &          0.3744           &          0.0902           &          0.0148           & 0.3762                    & 0.0895                    & 0.0150                    \\
			 CLASSY11   &          0.3730           &          0.0925           &          0.0142           & 0.3722                    & 0.0920                    & 0.0148                    \\
			Submodular  &          0.3888           &          0.0930           &          0.0141           & 0.3918                    & 0.0935                    & 0.0139                    \\
			    DPP     &          0.3992           &          0.0958           &          0.0159           & 0.3979                    & 0.0962                    & 0.0157                    \\
			  RegSum    &          0.3840           &          0.0980           &          0.0165           & 0.3857                    & 0.0975                    & 0.0160                    \\
			 OCCAMS\_V  &          0.3852           &          0.0976           &          0.0142           & 0.3850                    & 0.0976                    & 0.0133                    \\
			 ICSISumm   &          0.3855           &          0.0977           &          0.0185           & 0.3840                    & 0.0978                    & 0.0173                    \\ \hline
			Borda Count &          0.3700           &          0.0738           &          0.0115           & 0.3772                    & 0.0734                    & 0.0110                    \\
			    WCS     &          0.3815           &          0.0907           &          0.0120           & 0.3800                    & 0.0923                    & 0.0125                    \\
			 SentRank   &          0.3880           &          0.1010           &          0.0163           & 0.3870                    & 0.1008                    & 0.0159                    \\
			GlobalRank  &          0.3562           &          0.1045           &          0.0185           & 0.3955                    & 0.1039                    & \textbf{0.0191}$^\dagger$ \\
			 LocalRank  &          0.3992           &          0.1058           &          0.0192           & 0.3998                    & 0.1050                    & 0.0187                    \\
			HybridRank  & \textbf{0.4082}$^\dagger$ & \textbf{0.1102}$^\dagger$ & \textbf{0.0195}$^\dagger$ & \textbf{0.4127}$^\dagger$ & \textbf{0.1098}$^\dagger$ & 0.0180                    \\ \hline
		\end{tabular}
		\caption{Results on DUC 2003 dataset}
	\end{table*}
	
	\vspace{3mm}
	
	As shown in the table, the proposed systems outperform most existing systems on all three ROUGE scores. Both Borda and WCS failed to outperform the best state of art results. Even the simplistic SentRank algorithm outperforming most candidate systems and achieves a performance at par with the state of art systems in terms of ROUGE-2. While HybridRank achieves the best performance for ROUGE-1 and ROUGE-2 on both DUC 2003 and DUC 2004 datasets, GlobalRank performs best in terms of ROUGE-4 on DUC 2004. We performed a two sided sign test for determining whether the results were significantly different. The results clearly show that a rank aggregation technique that takes into account content of the summaries achieve a much higher ROUGE score vis-a-vis the systems that use only the ranked lists of sentences.
	
	\section{Conclusion}
	
	In this work we propose a new technique to estimate \emph{reliability} of a summarization system for a given document cluster. We define three systems, \emph{SentRank}, \emph{LocalRank} and \emph{GlobalRank} which take into account \emph{informativeness} of individual sentences, performance of candidates on a given document cluster, and overall performance of candidates on a held out development set, respectively. We use content overlap between summaries generated from several systems to estimate the relative importance of each system in case of LocalRank and SentRank. We combine the information from all these three systems to generate the final hybrid ranking (HybridRank) system. Summaries generated from such an aggregate system outperforms all the baseline and state of art systems as well as the baseline aggregation techniques by a good margin. 
	
	\bibliography{coling2018}
	\bibliographystyle{acl}
	
\end{document}